\documentstyle [12pt,epsfig]{article} 
\makeatletter
\@addtoreset{equation}{section}
\makeatother

\newcommand{\ba}{\begin{eqnarray}}
\newcommand{\ea}{\end{eqnarray}}

\begin{document}
\newcommand{\BS}{\bigskip}
\newcommand{\SECTION}[1]{\BS{\large\section{\bf #1}}}
\newcommand{\SUBSECTION}[1]{\BS{\large\subsection{\bf #1}}}
\newcommand{\SUBSUBSECTION}[1]{\BS{\large\subsubsection{\bf #1}}}

\begin{titlepage}
%
\vspace*{1cm}
\begin{center}
\vspace*{2cm}
{\large \bf On the Real and Apparent Positions of Moving Objects in Special
 Relativity: The Rockets-and-String and Pole-and-Barn Paradoxes Revisited
 and a New Paradox}  
\vspace*{1.5cm}
\end{center}
\begin{center}
{\bf J.H.Field }
\end{center}
\begin{center}
{ D\'{e}partement de Physique Nucl\'{e}aire et Corpusculaire
 Universit\'{e} de Gen\`{e}ve . 24, quai Ernest-Ansermet
 CH-1211 Gen\`{e}ve 4.
}
\newline
\newline
 E-mail: john.field@cern.ch 
\end{center}
\vspace*{2cm}
\begin{abstract}
  The distinction between the real positions of moving objects in a single reference
 frame and the apparent positions of objects at rest in one inertial frame and viewed
 from another, as predicted by the space-time Lorentz Transformations, is discussed.
 It is found that in the Rockets-and-String paradox the string remains unstressed and
 does not break and that the pole in the Barn-and-Pole paradox never actually fits into
 the barn. The close relationship of the Lorentz-Fitzgerald Contraction and the relativity
 of simultaneity of Special Relativity is pointed out and an associated paradox, in which
 causality is apparently violated, is noted
.

\vspace*{1cm} 
\end{abstract}
\end{titlepage}
 
\SECTION{\bf{Introduction}}
  One of the most important new physical insights given in Einstein's seminal
 paper on Special Relativity (SR)~\cite{Einstein} was the realisation that the
 Lorentz-Fitzgerald Contraction (LFC), which had previously been interpreted
 by Lorentz and Poincar\'{e} as an electrodynamical effect, was most easily
 understood as a simple consequence of the space-time Lorentz Transformation (LT),
 i.e. as a geometrical effect~\cite{Pais}. 
 \par It was pointed out 54 years later by Terrell~\cite{Terrell} and
 Penrose~\cite{Penrose} that when other important physical effects (light propagation
  time delays and optical aberration) are taken into account, as well as the LT,
  the moving sphere considered by Einstein in the 1905 SR paper would not
  appear to be flattened, in the direction of motion, into an ellipsoid,
  as suggested by Einstein, but rather would appear undistorted, but rotated.
  Shortly afterwards, Weinstein~\cite{Weinstein} pointed out that the LFC
  of a moving rod is apparent only if it is viewed in a direction strictly
  perpendicular to its direction of motion. It appears instead to be relatively
  elongated if moving towards the observer, and to be more contracted than the
  LFC effect if moving away from him. These effects are a consequence of
  light propagation time delays. A review~\cite{JHF1} has discussed
  in some detail the combined effects of the LT, light propagation delays
  and optical aberration on the appearence of moving objects and clocks.

 \par Another paper by the present author~\cite{JHF2} pointed out that, in addition
  to the well-known LFC and Time Dilatation (TD) effects corresponding, respectively,
  to $t = constant$ and  $x' = constant$ projections of the LT, two other 
  apparent distortions of space-time may be considered: Time Contraction (TC) and
  Space Dilatation (SD) corresponding to the two remaining projections,
 $x = constant$ and  $t' = constant$, respectively, of the LT
  \footnote{The space and time coordinates: $x$, $t$; $x'$, $t'$ are measured in two
 inertial frames, S; S' in relative motion along their common $x$-axis}.
  \par The apparent nature  of spatial distortions resulting from the LT is made evident
  by comparing the LFC with SD. In the former case the moving object appears shorter,
   in the latter longer, than the length as observed in its proper frame. Similarly,
   in TD, a moving clock appears to run slower than a similar clock at rest, while in
    TC identical moving clocks observed at a fixed position appear to be running
    faster than a similar, stationary, clock. A dynamical explanation has been proposed
    ~\cite{Sorensen} for the LFC effect in the case of an extended object bound by
   electromagnetic forces, and of TD~\cite{Jefimenko} in the case of various
   `electromagnetic clocks'. No attempt has been made, to date, to find a dynamical
   explanation of the TC and SD effects. It is hard to see, in any case, how any
   `dynamical' explanation can be given for the LFC of the proper distance between
   two isolated material objects in a common inertial frame, whereas, as first
   shown by Einstein, this is a simple space-time geometric consequence of the LT.
   As will be discussed below, one important reason for the misinterpretation
   of the `Rockets-and String paradox'~\cite{DewBer} is the incorrect assumption
   that extended objects undergo a `dynamical' LFC that is different from that of
   the distance between two discrete objects.
   \par The `Rockets-and-String' and `Pole-and-Barn'~\cite{Dewan} as well as the
   similar `Man-and-Grid'~\cite{Rindler} and `Rod-and-Hole'~\cite{Shaw} paradoxes 
   have all been extensively used in text books on SR, for example in Taylor and
   Wheeler~\cite{TayWheel} and, more recently, by Tipler and
   Lewellyn~\cite{TipLewell}. In all cases the apparent LFC effects are correctly
   derived from the LT equations, but it seems to be nowhere realised that the
   `real' positions and size of moving objects, in the sense that will be 
    described below in Section 2, are different from the apparent ones derived
    from the LT. Indeed, the general assumption, in all of the papers and books
    just cited seems to be that there is no distinction between the real and 
    apparent sizes and positions of moving objects, or, equivalently, that the
    LFC is a `real' effect. The purpose of the present paper is to point out 
    that this is not the case and that the `real' and `apparent' (i.e., those
     calculated using the LT) positions of moving objects are, indeed, distinct.
     The origin of
    this confusion between `real' and `apparent' positions and sizes of moving
    objects is not clear to the author. There is nothing in Einstein's 1905 
    SR paper to suggest that the LFC should be considered as a `real' rather
    than an `apparent' effect. Specifically, Einstein wrote~\cite{PerrJeff}:
    \par {\tt Thus, whereas the X and Z dimensions of the sphere (and therefore
     every rigid body of no matter what form) do not appear modified by the motion,
    \newline the X dimension appears shortened in the ratio $1 : \sqrt{1-v^2/c^2}$,
     i.e. the \newline greater the value of $v$ the greater the shortening.}
     \par In the original German the crucial phrase is: `erscheint die X-Dimension im
      Verh\"{a}ltnis $1 : \sqrt{1-v^2/c^2}...~$'. The verb `erscheinen', translated
      into English~\cite{GermDict} means `to appear'. Einstein never stated, in 
       Reference~\cite{PerrJeff}, that the LFC is a `real' effect.
     \par The plan of this paper is as follows: The following Section discusses 
      the real positions of objects as observed or measured in a single reference
      frame. Section 3 describes the apparent spatial positions of objects in
      uniform motion as predicted by the LT. The Rockets-and-String and Barn-and-Pole
      paradoxes are discussed in Sections 4 and 5 respectively. In Section 6, the
      close connection between the LFC and the relativity of simultaneity of SR is
      described and a new paradox of SR is pointed out. Section 7 contains a summary
      and conclusions.

\SECTION{\bf{The Real Positions of Moving Objects}}

\begin{figure}[htbp]
\begin{center}\hspace*{-0.5cm}\mbox{
\epsfysize10.0cm\epsffile{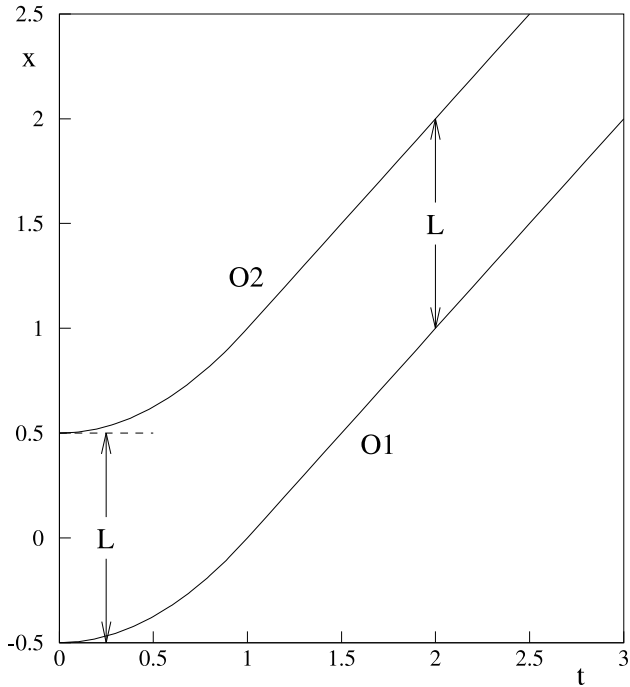}}
\caption{{\sl Space-time trajectories in the frame S of the real positions of O1 and O2
    when subjected to constant and identical proper accelerations.
  Units are chosen with $c  = 1$. Also $L= a = 1$, $t_{acc}=\sqrt{3}$.}}
\label{fig-fig1}
\end{center}
\end{figure}
 
 The `real' positions of one or more moving objects are defined here as those 
 specified, for the different objects, in a single frame of reference. The latter
 may be either inertial, or with an arbitary accelerated motion. By introducing
 the concept of a co-moving inertial frame, at any instant of the accelerated motion,
 the distance between the objects can always be defined as the proper distance 
 between them in a certain inertial frame. If this distance in the different co-moving
 inertial frames is constant the `real' distance is said to be constant. No distinction
 is made between the real distance between the points on a rigid extended object
 and that between discrete, physically separated, objects coincident in space-time with these
 points. This is because the LT, that relates only space-time events in different
 inertial frames, treats, in an identical manner, points on extended or discrete
 physical objects. 
 \par The utility of the science of the `real' positions of moving objects
  (astronomy, railways, military ballistics, air traffic control, space travel,
   GPS satellites...) is evident and the validity of the basic physical concepts
   introduced by Galileo (distance, time, velocity and acceleration) are not affected, in
   any way, by SR. The `real' positions of objects in a given reference frame are
   those which must be known to, for example, avoid collisions between moving
   objects in the case of railway networks, or, on the contrary, to assure them
   in the case of military ballistics or space-travel. From now on, in this paper,
   the words `real' and `apparent' will be written without quotation marks.
  \par For clarity, a definite measuring procedure to establish the real distance
   between two moving objects in a given frame of observation is introduced. This
   will be useful later when discussing the Barn-and-Pole paradox. Suppose that the 
    objects considered move along the positive x-axis of an inertial coordinate system S.
    The co-moving inertial frame of the objects is denoted by S'. It is imagined that
    two parallel light beams cross the x-axis in S, at right angles, at a distance
    $\ell$ apart. Each light beam is viewed by a photo-cell and the moving objects
    are equipped with small opaque screens that block the light beams during the
    passage of the objects. Two objects, O1 and O2, with x(O2) $>$ x(O1), moving with
    the same uniform velocity, $v$,  will
    then interrupt, in turn, each of the light beams. Suppose that the photo-cells
    in the forward (F) and backward (B) beams are equipped with clocks that measure
   the times of extinction of the beams to be (in an obvious notation) $t(F1)$,
    $t(F2)$, $t(B1)$ and $t(B2)$\footnote{`forward' and `backward' are  defined from the
    viewpoint of the moving objects. Thus the forward beam lies nearest to the origin 
    of the x-axis}. Consideration of the motion of the objects past the beams gives the 
    following equations;
    \begin{eqnarray}
     t(B1) -t(F1) & = & \frac{\ell}{v} \\
     t(B2) -t(F2) & = & \frac{\ell}{v} \\
     t(B2) -t(F1) & = & \frac{\ell-L}{v}
     \end{eqnarray}
     where $L$ is the real distance between the moving objects.
      Eqns(2.1) and (2.2) give the times of passage of the objects O1 and
      O2, respectively, between the light beams, whereas Eqn(2.3) is 
      obtained by noting that, if $\ell > L$, each object moves the distance $\ell-L$
      during the interval $t(B2)-t(F1)$. If  $\ell < L$ 
      each object moves a distance $L-\ell$ during the interval $t(F1)-t(B2)$ and
      the same equation is obtained. Taking the ratios of Eqn(2.3)
      to either Eqn(2.1) or Eqn(2.2) yields, after some simple algebra, the
      relations: 
      \begin{eqnarray}
      L  & = & \ell \frac{[t(B1)-t(B2)]}{t(B1)-t(F1)} \\
      L  & = & \ell \frac{[t(F1)-t(F2)]}{t(B2)-t(F2)}
     \end{eqnarray}
      Subtracting Eqn(2.3) from Eqs(2.1) or (2.2), respectively, gives:
     \begin{eqnarray}
     t(B1) -t(B2) & = & \frac{L}{v} \\
     t(F1) -t(F2) & = & \frac{L}{v}
     \end{eqnarray}
      Taking the ratios of Eqn(2.6) to (2.2) and Eqn(2.7) to (2.1)
      then yields two further equations, similar to (2.4) and (2.5) above:
     \begin{eqnarray}
      L  & = & \ell \frac{[t(B1)-t(B2)]}{t(B2)-t(F2)} \\
      L  & = & \ell \frac{[t(F1)-t(F2)]}{t(B1)-t(F1)}
     \end{eqnarray}
      Equations (2.4),(2.5),(2.8) and (2.9) show that any three of the
      four time measurements are sufficient to determine the real separation, $L$, 
      between the two moving objects. In these equations the times: $t(F2)$, $t(B1)$, 
     $t(F1)$ and $t(B2)$, respectively, are not used to determine $L$. In order
      to combine all four time measurements to obtain the best, unbiased, determinations of
      $v$ and $L$, Eqns(2.1) and (2.2) may be added to obtain: 
      \begin{equation}
       t(B1)+t(B2)-t(F1)-t(F2) = \frac{2 \ell}{v}
      \end{equation}
       while subtracting two times Eqn(2.3) from (2.10) gives:
        \begin{equation}
       t(B1)-t(B2)+t(F1)-t(F2) = \frac{2 L}{v}
      \end{equation}
     The velocity, $v$, is obtained by transposing Eqn(2.10):
         \begin{equation}
     v = \frac{2\ell}{t(B1)+t(B2)-t(F1)-t(F2)}
    \end{equation}
     while the ratio of Eqn(2.11) to (2.10) gives:
    \begin{equation}
     L = \ell \frac{[t(B1)-t(B2)+t(F1)-t(F2)]}{t(B1)+t(B2)-t(F1)-t(F2)}
    \end{equation} 
     \par It is interesting to note that the
     real spatial positions, as well as the instantaneous velocity and acceleration, at
     any time, of two objects subjected to a symmetric, uniform, acceleration
     in the frame S, can also be determined from the four time measurements
     just considered. In this case there is no redundancy, the time measurements
     determine four equations which may be solved for the four quantites
     just mentioned. In the case of uniform motion, the constant velocity 
     hypothesis may be checked by comparing the independent determinations
      of $v$ provided by Eqns(2.1) and (2.2). Furthermore, as already
      mentioned, any three of
     the four time measurements is sufficient to determine $L$.
     Evidently, if  $t(F1) = t(B2)$, then $L = \ell$ in the case of an  
     arbitary accelerated motion of the two objects. Thus, by varying  $\ell$, the real
    distance between the co-moving objects can be determined even if the 
    acceleration program of their co-moving frame is not known.
     \par The two objects, moving with equal and constant velocities along the x-axis
     in S, discussed above, are now considered to be set in motion by applying 
     identical acceleration programs to two objects initially at rest and lying
       along the $x$-axis in S.
  The two objects considered are then, by definition,
  subjected to the same acceleration program in their common rest frame, or, what is the 
  same thing, their common rest frame, (with respect to which the two objects are, at all times,
   at rest) is accelerated. Under these circumstances the distance between the objects 
  remains constant in the instantaneous co-moving inertial frame of the objects. At the
   start of the acceleration procedure, the instantaneous co-moving inertial frame is S,
   at the end of the acceleration procedure it is S'. Therefore the separation of the
   objects in S at the start of the acceleration
   procedure is the same as that in  S' at the end of it. Note that there is no distinction between the
  real and apparent distances for objects at rest in the same inertial frame. Also `relativity
  of simultaneity' can play no role, since the proper time of both objects is always referred to the 
   same co-moving inertial frame. Since the acceleration program of both objects starts
   at the same time in S, and both objects execute identical space time trajectories, the real
    separation of the objects must
   also remain constant. This necessarily follows from space-time geometry. Similarly, since the 
   acceleration program stops at the same time in S' for both objects the real separation of the objects 
   remains constant in this frame and equal to the original separation of the objects
   in S. This behaviour occurs for any symmetric acceleration program, and is shown,
    for the special case of a constant acceleration in the rest frame of the objects (to be calculated
   in detail below), in Fig 1.
   Since, in the above discussion, both objects are always referred to the same inertial
    frame there is no way that SR can enter into the discussion and change any of the above conclusions.
   Indeed, SR is necessary to derive the correct form of the separate space-time trajectories
   in S, but the symmetry
   properties that guarantee the equalites of the real separations of the objects cannot be 
   affected, in any way, by SR effects. 

    \par Two objects, O1 and O2, originally lying at rest along the x-axis in
     S and separated by a distance $L$ are now simultaneously accelerated, 
     during a fixed time period, $t_{acc}$, in S, starting at $t=0$,
     with constant acceleration, $a$, in their common proper frame, up to a 
     relativistic velocity $v/c \equiv \beta = \sqrt{3}/2$,  corresponding to
     $\gamma = 1/\sqrt{1-\beta^2} = 2$, and $t_{acc} = c\sqrt{3}/a$ . The equations giving the velocity
     $v$ and the position $x$ in a fixed inertial frame, using such an
     acceleration program, were derived by
      Marder~\cite{Marder} and more recently by Nikolic~\cite{Nikolic} and Rindler~\cite{Rindler1}.    
      The positions and velocities of the
      objects in the frame S are:
      \newline
      \newline for $t \le 0$
       \begin{eqnarray}
         v_1 & = & v_2 = 0  \\
         x_1 & = & -\frac{L}{2} \\
         x_2 & = & \frac{L}{2}
       \end{eqnarray}
        for $0 < t < t_{acc}$
  \begin{eqnarray}
     v_1(t) & = & v_2(t) = v(t)  =  \frac{act}{\sqrt{c^2+a^2t^2}}   \\
      x_1(t) & = & c\left[\frac{\sqrt{c^2+a^2t^2}-c}{a}\right]-\frac{L}{2} \\   
      x_2(t) & = & c\left[\frac{\sqrt{c^2+a^2t^2}-c}{a}\right]+\frac{L}{2} 
    \end{eqnarray}        
    and for $t \ge t_{acc}$
  \begin{eqnarray}
     v_1(t) & = & v_2(t) = v(t_{acc})   \\
      x_1(t) & = &  v(t_{acc})(t-t_{acc})+x_1(t_{acc}) \\
      x_2(t) & = &  v(t_{acc})(t-t_{acc})+x_2(t_{acc})
    \end{eqnarray} 
 
    The origins of S and S' have been chosen to coincide at $t = t' =0$.
    The real positions of the objects in S are shown, as a function of 
    $t$ for $a=1$ and $t_{acc}=\sqrt{3}$, in Fig.1. 
     The velocities of the two objects
    are equal at all times, as is also the real separation of
    the objects $x_2-x_1 = L$. 
    SR is used only to derive Eqn(2.17). The time-varying velocity
    is then integrated according to the usual rules of classical
    dynamics in order to obtain Eqns(2.18),(2.19) for the 
    positions of the objects during acceleration. These are, by definition,
    the {\it real} positions of the objects O1 and O2 in S.
     The equalities of the
    velocities and the constant real separations are a direct consequence of
    the assumed initial conditions and the similarity of the proper frame 
    accelerations of the objects. These are the sets of equations that must
    be used to specify the distance between O1 and O2 and any other objects,
    whose real positions are specified in S, in order to predict
    collisions or other space-time interactions of the objects. 

\SECTION{\bf{The Apparent Positions of Moving Objects in Special Relativity}}
 
    In order to account correctly for the actual appearence of moving objects,
   not only the LT but also the effects of light propagation delays to the observer, 
  and optical aberration must be properly taken into account. In the following, 
   it is supposed that the necessary corrections for the two latter effects have
   been made, so that only consequences of the LT are discussed. In this case the
   apparent position $x^A$ in a `stationary' inertial frame, S, at
    time\footnote{Notice that there
   is no possible distinction between the `time' and the `apparent time'
   in the frame S. $t$ is simply the time recorded by a synchronised local clock in S.},
   $t$, 
   of an event at $x'$, $t'$ in the inertial frame, S', moving with 
    velocity $v$ along the positive $x$-axis in S, is given
   by the space-time LT:
   \begin{eqnarray}
   x^A & = & \gamma (x'+v t')   \\
   t & = & \gamma (t'+\frac{vx'}{c^2})
  \end{eqnarray}
   In Eqns(3.1) and (3.2), clocks in S and S' synchronised so as to
   record the same time at $t = t'= 0$ when  $x^A = x'= 0$. 
   \par Since `observation'
   implies the interaction, with a detection 
   apparatus, of one or more photons, originating from the
    observed object, it is convenient to define the space-time events that are related
   by the LT to correspond, in S', to the points of emission (or reflection) of a 
   photon from a source at rest, and in S to the space-time point at
   which the emission or reflection of this photon is observed. It is assumed that,
    after the acceleration
   procedure described in the previous Section, the frame S' moves with 
   constant velocity $ v = V$ along the x-axis of S. How the object will appear
   when viewed from S, depends on which photons, scattered from or emitted by
   the object, are selected by the observer. In the case of the LFC discussed
    by Einstein~\cite{Einstein} the observer in S requires that the photons 
    emitted from O1 and O2 are observed simultaneously in S (the $\Delta t = 0$
    projection of the LT). As is easily derived from Eqns(3.1) and (3.2), the
     separation of the two objects
    then appears to be $L/\gamma$ where $L$ is their real separation in S'.
    Another simple possibility~\cite{JHF2} is to require that the two objects
    are both illuminated for a very short period of time in their proper
     frame, yielding a `transient luminous object'. In this case, a quite
    complicated series of events is observed in S. A line image (whose width
     depends on the period of illumination) is seen to move with velocity
     $c/\beta$ in the same direction as that of the moving object, sweeping out
     a total length $\gamma L$. This is the so-called `Space Dilatation' (SD)
     effect (the $\Delta t' = 0$ projection of the LT)~\cite{JHF2}.
     As discussed in more detail below, both the LFC and TD effects are direct
     consequences of the `relativity of simultaneity' of SR introduced by Einstein.
     In both LFC and TD, the space-time events in S' are associated with the 
     real positions of the objects, as defined above, in this frame. However,
     since it is clear that the apparent positions of these space-time events, 
     as observed in S, depend upon the mode of observation, being different for the
     LFC and TD, they evidently cannot be identified with the real positions
     of the objects in this frame. The relation of the real and apparent positions
     will now be discussed in detail for the two `paradoxes' of SR mentioned above.

\SECTION{\bf{The Rockets-and-String Paradox}}

    The idea that the LFC could induce mechanical stress in a moving extended
   body was introduced by Dewan and Beran~\cite{DewBer}, criticised by
   Nawrocki~\cite{Nawrocki} and defended by Dewan~\cite{Dewan} and 
   Romain~\cite{Romain}. Dewan and Beran discussed symmetric acceleration
   of two rockets, originally at rest in the frame S, in a way very similar to the
   acceleration of the two objects O1 and O2 described in Section 2 above, except
   that no specific acceleration program was defined. It was correctly concluded
   that the real distance between the objects in S would remain unchanged 
   throughout the aceleration procedure. However, it was not stated
   that, after acceleration, the proper separation between the rockets is the 
   same as their original separation in S. Dewan and Beran then introduced a 
   continous string attached between the rockets during the acceleration and
   drew a distinction between two distances:
   \par{\tt (a) the distance between two ends of a connected rod and (b)
     the \newline distance between two objects which are not connected but each of which 
      \newline independently and simultaneously moves with the same velocity with respect
      \newline to an inertial frame}
   \par It was then stated (without any supporting argument) that the distance
      (a) is subject to the LFC and (b) not. Replacing the `connected rod' of
     (a) by a continous string attached between the rockets, it was concluded 
     that the string would be stressed and ultimately break, since the distance
     between the rockets does not change, whereas the string shrinks due to the
    LFC. Several correct arguments were given why the {\it real} distance
     between the rockets does not change. The incorrect conclusion that the
    string would be stressed and break was due to the failure to discriminate
    between the {\it real} separation of the rockets (correctly calculated)
    and the {\it apparent} contraction due to the LFC, which as correctly 
    pointed out by  Nawrocki~\cite{Nawrocki} applies equally to the distance
    between the ends of the string and that between the points on the 
    rockets to which it is attached. Dewan and Beran's distinction
    between the distances (a) and (b) is then wrong. Both the distance
    between the points of attachment of the string and the length of the string
    undergo the same apparent LFC. There is no stress in the string. It does not
    break. 
    \par Dewan did not respond to Nawrocki's objection that (correctly) stated 
    the equivalence of the length of an extended object and the distance 
    between two independant objects separated by a distance equal to this
    length, but instead introduced a new argument, claiming that Nawrocki
     had not correctly taken into account the relativity of simultaneity of SR.
    Dewan considered the sequence of events corresponding to the firing of
    the rockets as observed in S', the co-moving frame of the rockets after
    acceleration. It was concluded that the spring breaks because, in this case,
    the final separation of the rockets is $\gamma L$ (this is just the SD effect
    described above). But for this, the LT must be applied to space time events
     on the rockets (and not to the string), whereas when considering an observer
     in S, Dewan and Beran had applied the LT to the string (and not to the
     rockets)! Viewed from S' both the distance between the rockets and the
      length of the string undergo the apparent SD effect, there is no 
      stress in the string and it does not break. 
     \par All of the confusion and wrong conclusions in References~\cite{DewBer}
      and~\cite{Dewan} result, firstly, from the failure to notice the strict 
       equivalence of separations of type (a) and (b) and secondly from not
      realising that the apparent and real separations of the moving rockets
     are not the same. Even Nawrocki, who correctly pointed out the fallacy
        of assigning the LFC to separations of type (a) but not to type (b)
     seems not to have realised the correctness of Dewan and Beran's calculation
    of the real distance between the rockets. He correctly stated that the 
    apparent contractions of the length of the string and the distance 
    between the rockets would be the same, but did not discriminate this
   apparent distance from the real separation of the rockets in the frame S.
    \par The acceleration program discussed in Section 2 above (and indeed 
     any programme which is an arbitary function of proper time) when applied
    in a symmetrical manner to the two objects O1 and O2, will leave unchanged
    the proper separation of the objects in their co-moving frame. On the 
    other hand. the `instantaneous' acceleration considered by Dewan 
    apparently resulted in a change in the proper separation of the rockets. 
    If Dewan had considered instead a physically realistic acceleration
    program, such as that discussed in Section 2 above (which must necessarily
    occupy a finite period of time) no change in the proper separation of 
    the rockets in their co-moving frame would have been predicted. This is shown in
     Fig.1b above. In fact, Dewan did not consider the whole acceleration
    process, but instead calculated the apparent positions in S', according 
    to the LT, of the space-time events corresponding to the firing of the 
    rockets in S, i.e. the start of the acceleration programme. These apparent
    positions are not the same as the real positions of the rockets in S'.
    The apparent length of a rod uniformly accelerated in its proper co-moving
     frame has recently been calculated by Nikolic~\cite{Nikolic}. It was 
     correctly noted in this paper, contrary to the conclusion of Dewan, that the real
    length of the rod remains unchanged in this frame.
    \par In a comment on Reference~\cite{DewBer}, Evett and Wagness~\cite{EvWag}
    stated that the distance between the rockets, given by Dewan and Beran as that
    between the tail of the first rocket and the head of the second, would  not
    remain constant in S, due to the LFC of the rockets, whereas the distance 
    between any two corresponding points would, Thus, like Dewan and Beran, 
    Evett and Wagness assumed that extended material objects (rockets, strings)
    undergo the LFC, but not the spatial separation of the rockets in their
    co-moving frame. Thus  Evett and Wagness also stated that the string will
     break. Actually, the real lengths of the rockets as well as the distance
     between them, in the frame S, remain unchanged during acceleration, but, as
     correctly stated by Nawrocki, they both undergo the apparent LFC.
    \par In Section 3 of Reference~\cite{Dewan} Dewan considered an observer
     at rest in the co-moving frame of the accelerated rocket, R2, with the
    smallest x-coordinate. He concluded that, viewed from this frame, the
    second rocket, R1, would have a non-vanishing velocity in the x-direction
    that stresses the string and causes it to break. Such behaviour is
    clearly in contradiction with the assumed symmetry of the acceleration
    program. At any instant the two rockets have a common co-moving frame
    within which their relative velocity is zero. Dewan's argument was later
   supported by Romain~\cite{Romain} who considered a constant acceleration of the 
   rockets in their co-moving proper frame as in Eqns.(2.17)-(2.19) above.
   This conclusion however was founded on a misinterpretation of the space-time
   diagram (Fig.1 of Reference~\cite{Romain}) describing the motion of the rockets.
   In reference to this figure it was incorrectly stated that `$B_2B_1'$ and $A_2A_1$
   represent the same ``proper length'' '. In fact, the proper separation of the
   rockets in the frame S' (with space and time coordinates $X'^1$ and $X'^4$)
    corresponding to the space-time point $B_2$, on the trajectory of R2,
    in the frame S (with space and time coordinates
    $X^1$ and $X^4$)
   is represented by the distance between $B_2$ and the intersection of
   the $X'^1$ axis with the tangent to the trajectory of R1 at $B_1$ (say, 
    the point $B_1^t$) not $B_2B_1'$. The latter is the distance between $B_2$
     and the intersection of the $X'^1$ axis with the trajectory of R1.
    The apparent distance in S corresponding to the segment $B_2B_1^t$
    (a $\Delta X'^4 = 0$ projection) is found to be
    $\gamma B_2 B_1$ so that, from the SD effect described in Section 3
    above, the proper separation of the rockets in S' is $B_2B_1 = A_2A_1$, i.e.
    the same as their initial separation in S. Thus it is the segment $B_2B_1^t$,
    not $B_2B_1'$, that represents the same proper length as $A_2A_1$. Since the
    velocity in S of R1 at $B_1'$ is greater than that of $R_2$ at $B_2$, Romain
    was lead to the same erroneous conclusion as Dewan, that the proper distance
    between the rockets is increasing, causing the string to break. It may be noted
    that this conclusion is actually in contradiction with Romain's own (but incorrect)
    statement that $A_2 A_1$ and $B_2B_1'$  represent the same proper length! Although
     Romain did 
    not discriminate between the real and apparent positions of the rockets, it may
    be remarked that the trajectories $A_2C_2$, $A_1C_1$ in Fig.1 of
    Reference~\cite{Romain} actually represent the real positions of the rockets in S as
    discussed in Section 2 above. This figure is thus similar to Fig.1a of the present
    paper. Also the distance $B_2B_1^t$ represents, in S', the constant (real) separation
    of the rockets R1, R2 in their common co-moving inertial frame. 
    \par The apparent positions of the similarly accelerated rockets were later
     discussed by Evett~\cite{Evett}. No distinction was made between the real
     and apparent  positions of the rockets and, unlike Romain, no definite 
     acceleration procedure was defined. Like Dewan, a misinterpretation of the
     space time events corresponding to the firing of the rockets, as observed
     from their final inertial frame, lead to the erroneous conclusion that the
     real separation of the rockets in this frame is different from the initial
     one before acceleration.
   \par What is essentially the `Rockets-and-String Paradox' has been
   recently re-discussed by Tartaglia and Ruggiero~\cite{TR}, although no
   reference was made, in their paper, to the previous literature on the 
   subject discussed in the present paper. As in Section 2 above, the case of two objects
   (spatially independent or connected by a spring), subjected to an idential 
   uniform acceleration in their proper frames, as in Section 2 above, 
   was considered, as well as 
   that of an extended object (rod) subjected to a similar acceleration
    program. No distinction was made between the real and apparent 
    positions of the objects considered, and no calculations using the space-time
    LT were performed. Exactly the same error as that of Dewan and Beran was made
    by introducing an incorrect distinction between the spatial interval between
     the ends of a connected rod and that between two 
    spatially disconnected objects situated at the same spatial positions
     as the ends of the rod.
    Thus, it was concluded that the accelerated rod would be length contracted
    by an amount determined by its instantaneous velocity in S, as well as
    undergoing mechanical
   deformation due to elastic stress generated by the accelarating force.
   No calculations were performed, or arguments given, to justify this 
   conclusion; it is simply stated that:
  \par{\tt The length seen by O is obtained from
   $l'$ projectiong it from the x' axis (the space of O') to the x axis,
   i.e. multiplying by $\sqrt{1-v^2/c^2}$, according to the standard Lorentz contraction.}
   \par Thus here, Tartaglia and Ruggiero were referring to the apparent, not the
   real, length of the rod. In the case of two independent, and similarly
   accelerated, objects, the real trajectories in S were correctly described,
   as in Eqns(2.14)-(2.22) above, and it was correctly stated that the 
   separation of the objects remains constant in this frame. The real separation 
   was, however, confused with the apparent separation between the objects, since
   it was then stated, again, without any supporting argument or calculation,
   that:
 \par{\tt .....the proper distance in the frame of the rockets, $l_0$, has \newline
    progressively increased during the acceleration so that the Lorentz \newline
    contraction ($l =\gamma^{-1}l_0$) produces precisely the $l$ result.}
    \par The general arguments and detailed calculations presented in Section 2
    above clearly show that this statement is incorrect. There is no change in the
    separation of the objects in their co-moving inertial frame during the
    acceleration procedure. In the case when the two objects are connected
     by a spring, Tartaglia and Ruggiero incorrectly argued that the spring,
    being an extended object, undergoes Lorentz contraction whereas the spatial
    separation between the objects does not, thus inducing tension in the
     spring. It is then argued that:
 \par{\tt The consequence will be that the actual acceleration of the front
  end \newline will be a little bit less than what the engine alone would produce
  and the \newline acceleration at the rear end will be a little bit more for the
   same 
   reason. In this way, the proper times of the two engineers will no
   longer be the \newline same at a given coordinate time and the two world lines
  of the ends of the spring will no longer be equal hyperbolae (see figure 2).}
    \par Since, in fact, there is no distinction for the LT, between the space 
    interval between points on spatially separated objects, or an equal
   interval between points on a spatially extended object, there is no `relativistic
   tension', generated in the spring, to give rise to the different accelerations of
    the two objects as conjectured in Reference~\cite{TR}.
    Since the distances between
    the ends of the rod, the two spatially separated objects or
   the two connected objects remain always constant in the co-moving 
   inertial frame, during the acceleration procedure, it follows that
  in all three cases considered in Reference~\cite{TR}, the real distance, in S, between the
  moving objects (except for
  some possible elastic deformation in the case of the
   accelerated rod)  will
   be the same as that shown in Fig.1 of Reference~\cite{TR}. This figure is
   similar to Fig. 1a of the present paper or Fig. 1  of Reference~\cite{Romain}.
   Fig. 2 of Reference~\cite{TR} is therefore incorrect. Finally, it 
   may be noted that the apparent length of a uniformly
   accelerated rod has been calculated by Nikolic~\cite{Nikolic}. The 
   result obtained does not agree with the naive generalisation of the
   Lorentz contraction formula as given after Eqn(3) in Reference~\cite{TR}.

\SECTION{\bf{The Pole-and-Barn Paradox}}
\begin{figure}[htbp]
\begin{center}\hspace*{-0.5cm}\mbox{
\epsfysize10.0cm\epsffile{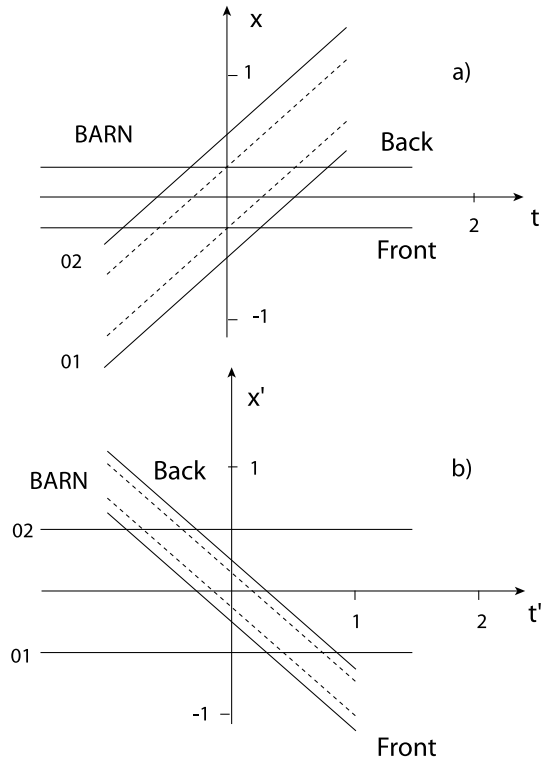}}
\caption{{\sl Space-time trajectories of the real (full lines)
 and apparent (dashed lines) positions of O1 and O2 in the region of the barn. a) in S, b) in S'.}}
\label{fig-fig2}
\end{center}
\end{figure}

\begin{figure}[htbp]
\begin{center}\hspace*{-0.5cm}\mbox{
\epsfysize10.0cm\epsffile{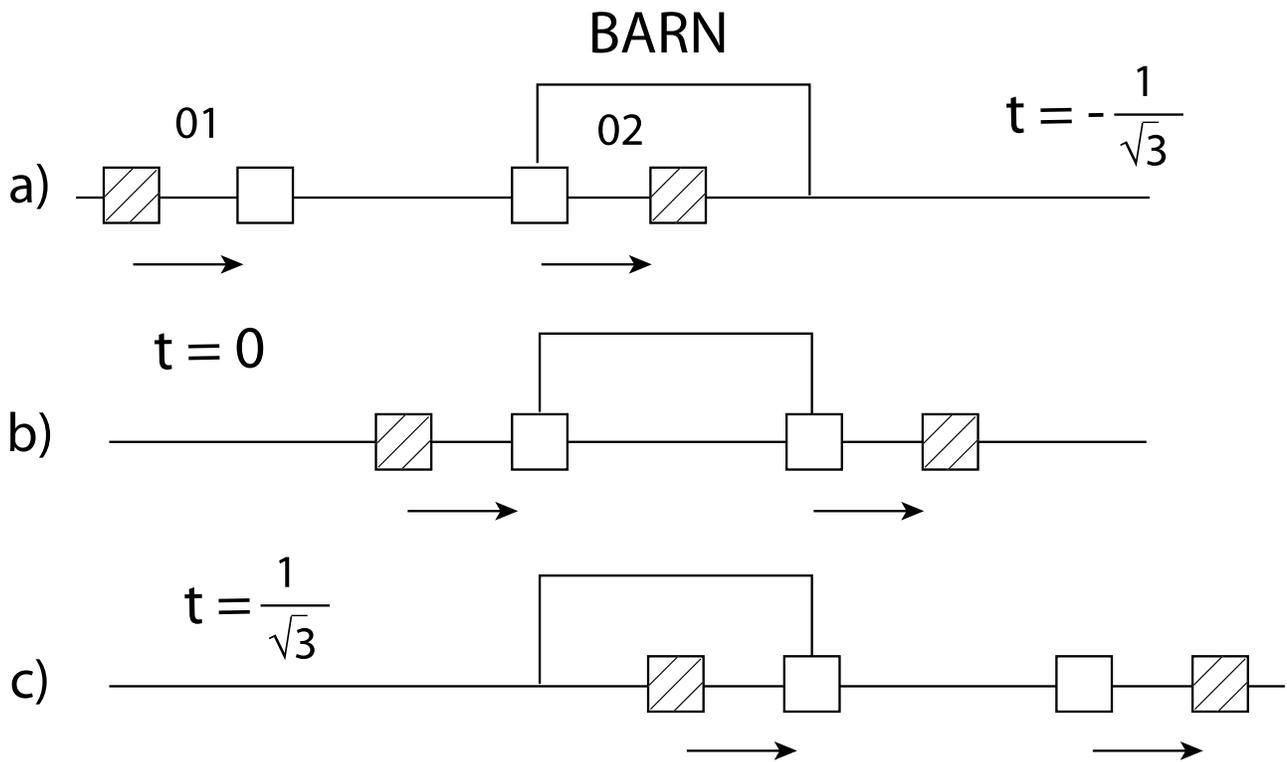}}
\caption{{\sl Real (cross-hatched squares) and apparent (open squares) positions of
  O1 and O2 in S, as they pass by the barn.}}
\label{fig-fig3}
\end{center}
\end{figure}

\begin{figure}[htbp]
\begin{center}\hspace*{-0.5cm}\mbox{
\epsfysize10.0cm\epsffile{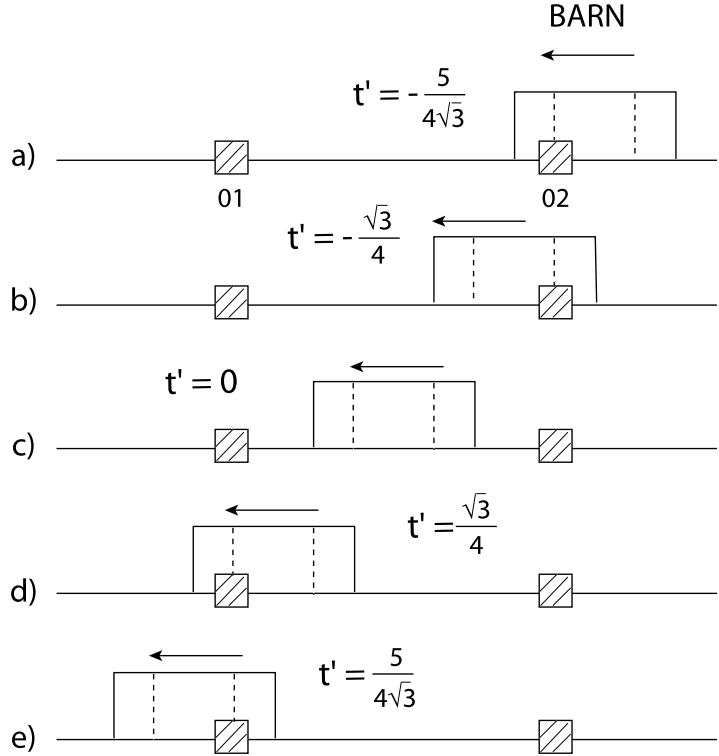}}
\caption{{\sl The moving barn as observed from S'. The real and apparent positions 
 of the ends of the barn are denoted by solid and dashed lines respectively.}}
\label{fig-fig4}
\end{center}
\end{figure}
\begin{table}
\begin{center}
\begin{tabular}{|c|c|c|} \hline
Time interval & O1 & O2 \\ \hline 
  $ t \le -\frac{1}{\sqrt{3}}$ & OUT & OUT \\ 
  $ -\Delta \ge t > -\frac{1}{\sqrt{3}}$ & OUT & IN \\ 
  $ \Delta > t >  -\Delta $ & IN & IN \\ 
  $ \frac{1}{\sqrt{3}} > t \ge \Delta$ & IN & OUT \\ 
  $ t \ge \frac{1}{\sqrt{3}}$ & OUT & OUT \\         
\hline
\end{tabular}
\caption[]{{ \sl Apparent positions (IN or OUT of the barn) of O1 and O2
   as viewed in S. The barn is assumed to be slightly
  longer than the Lorentz-contracted distance between O1 and O2, so
   that $\Delta \ll 1$.} } 
     
\end{center}
\end{table}
 
\begin{table}
\begin{center}
\begin{tabular}{|c|c|c|} \hline
Time interval & O1 & O2 \\ \hline 
  $ t' \le -\frac{5}{4\sqrt{3}}$ & OUT & OUT \\ 
  $  -\frac{\sqrt{3}}{4} \ge t' >  -\frac{5}{4\sqrt{3}}$ & OUT & IN \\ 
  $  \frac{\sqrt{3}}{4} > t' > -\frac{\sqrt{3}}{4}$ & OUT & OUT \\ 
  $ \frac{5}{4\sqrt{3}} > t' \ge \frac{\sqrt{3}}{4}$    & IN & OUT \\
  $ t' \ge  \frac{5}{4\sqrt{3}} $ & OUT & OUT \\         
\hline
\end{tabular}
\caption[]{{ \sl Apparent positions (IN or OUT of the barn) of O1 and O2
   as viewed in S'. The length of the barn is equal to the Lorentz
   contracted separation of the objects as viewed in S. }}     
\end{center}
\end{table} 

\begin{table}
\begin{center}
\begin{tabular}{|c|c|c|} \hline
Time interval & O1 & O2 \\ \hline 
  $ t \le -\frac{\sqrt{3}}{2}$ & OUT & OUT \\ 
  $  -\frac{1}{2\sqrt{3}} \ge t >-\frac{\sqrt{3}}{2}$ & OUT & IN \\
  $ \frac{1}{2\sqrt{3}} > t > -\frac{1}{2\sqrt{3}} $ & OUT & OUT \\ 
  $ \frac{\sqrt{3}}{2} > t \ge \frac{1}{2\sqrt{3}} $    & IN & OUT \\ 
  $ t \ge \frac{\sqrt{3}}{2} $ & OUT & OUT \\         
\hline
\end{tabular}
\caption[]{ { \sl Real positions (IN or OUT of the barn) of O1 and O2
   in S. The real positions in S' are given by the replacement $t \rightarrow t'$.
   The length of the barn is equal to the Lorentz
   contracted separation of the objects as viewed in S.}}     
\end{center}
\end{table} 
  
     For convenience, and without any loss of generality, the `pole' of the example
  is replaced by two independent objects O1 and O2, separated by unit distance,
  that are symmetrically accelerated from rest in the inertial frame S (in which the
  barn is at rest), by the procedure described in Section 2 above, to the inertial
  frame S', co-moving with the objects. Units are chosen such that $c =1$.
   Also, as in Fig.1, $L = a = 1$ and $t_{acc}= \sqrt{3}$. If
   the length of the barn is $0.5+\delta$, where $\delta$ is a small number,
    then, due to the LFC, the apparent positions of O1 and O2, viewed in S, will,
   at some time, both fit into the barn. This example was introduced by Dewan~\cite{Dewan},
   at the end of the paper written in reply to Nawrocki's objections to Dewan and Beran's
   `Rockets-and-String paradox' paper. Dewan also imagined that the doors of the barn
    might be closed when the pole was inside, and that if the pole was suddenly stopped
    by the closed back door of the barn, the rod would revert to its original proper
   length, thus crashing through the closed front door of the barn (Figs.1 and 2 of
    Reference~\cite{Dewan}). 
    \par In the associated `paradox' Dewan supposes that a pole vaulter is running along 
    carrying the pole:
    \par {\tt The pole vaulter, on the other hand, would see the barn as contracted
      \newline and much
      smaller than the pole. The question is, how can the pole
      \newline vaulter explain the fact
     that the door can be closed behind him?}
     \par All of the above takes for granted that the pole is really, not apparently,
       contracted in the frame S. However, the actual situation during the passage of
     the pole (i.e the objects O1 and O2) through the barn, is shown in Figs. 2, 3 and 4.
     In Figs.2a,b the real and apparent positions of O1 and O2, as a function of time,
     are shown in S and S' respectively. The clocks in S and S' are synchronised at the
     moment that the mid-point of O1 and O2 is situated at the middle of the barn. The real
     positions in S of O1 and O2, as calculated in Section 2 above, are indicated by the full
     lines, the apparent positions, due to the LFC, by the dashed ones. The real and
     apparent positions
     of the barn (at rest in S),  as viewed from S' are similarly
     indicated. It is quite clear
     that at $t = t' =0$, when the mid-point between O1 and O2 co-incides with that of the
     barn, O2 is already beyond the back door and O1 still before the front door. 
     At no time are both objects inside the barn at the same time, so that the doors
     cannot be closed with them both inside. The passage of the objects through the barn
     is shown in more detail in Fig.3 (as viewed in S) and in Fig.4 (as viewed in S').
     The real and apparent positions of the objects are shown as cross-hatched and open
     squares respectively. The times at which the objects apparently enter and leave the
     barn are shown in Table 1 for S and Table 2 for S'. Assuming that the barn is slightly
     longer than the Lorentz-contracted distance between the objects, it can be seen that,
     for a short period of time, both objects will appear to be inside the barn, as viewed
     from S, but that this is never the case for an observer in S'. Thus the apparent
     object `barn with the doors closed and both objects inside' exists for an observer
     in S, but not for one in S'. This might be possible for the apparent positions,
     but must evidently lead to a contradiction if the `barn with the doors closed
      and both objects inside' is associated with a real physical object, as was done
      by Dewan.

 \par The paradox was discussed in Reference~\cite{TayWheel} where it was pointed out
  that, due to the relativity of simultaneity, and as can be seen in Fig.4 and Table 2,
  for an observer in S', O2 will leave the barn at $t' = -\sqrt{3}/4$ (Fig.4b)
  before O1 enters it at $t' = \sqrt{3}/4$ (Fig.4d). If the (apparent) doors of
  barn close at these instants (and the barn is slightly longer than 0.5 units) then
  O2 will be inside the barn when the front door closes, and O1 will be inside when
  back door closes, as also seen by an observer in S. It was not pointed out, however,
  in Reference~\cite{TayWheel} that the observer in S', unlike the one in S, never
   sees both objects in the barn at the same instant. The times at which the moving objects
  really enter and leave the barn in S are shown in Table 3. The corresponding
  times in S' are simply given by the replacement $t \rightarrow t'$.
   \par In conclusion, the situation shown in Figs.1 and 2 of Reference~\cite{Dewan} is
    a physically impossible one. The pole can never, at any time, really fit into
    the barn. As in the case of the Rockets-and-String paradox it has been incorrectly
    assumed that the apparent Lorentz contracted length of the pole in S is equal
    to its real length (as defined in Section 2 above) in this frame.
   \par The real separation of the moving objects in S could be measured, as described
    in Section 2 above, by a parallel pair of light beams, perpendicular to the direction
    of motion of the objects, situated at the positions of the doors of the barn. Substituting
     the real times of passage of the objects given in Table 3 into Eqn.(2.13) gives their real
     separation: $L = 2 \ell = 1$ unit.

\SECTION{\bf{The Lorentz-Fitzgerald Contraction and the Relativity of Simultaneity}}

\begin{figure}[htbp]
\begin{center}\hspace*{-0.5cm}\mbox{
\epsfysize10.0cm\epsffile{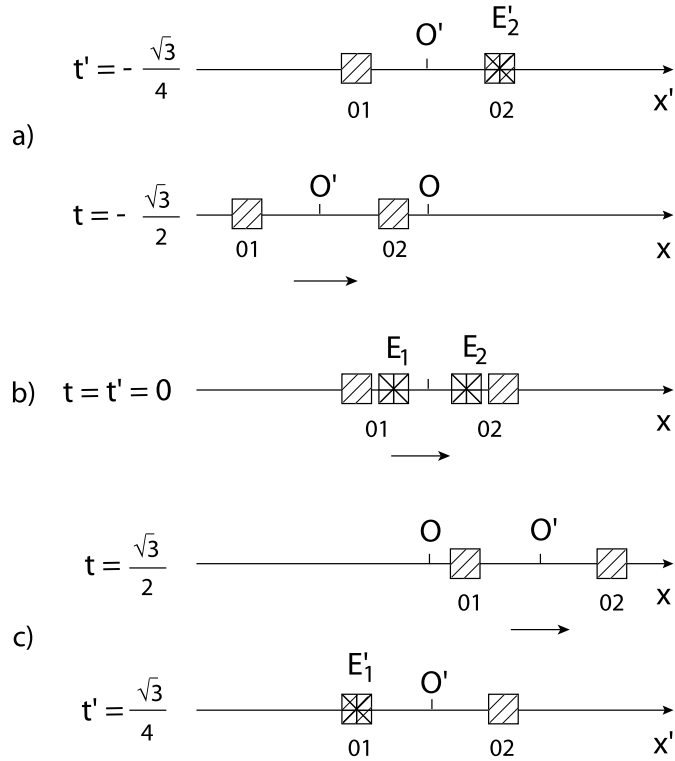}}
\caption{{\sl Space-time events $E_1'$ and $E_2'$ in S' and the real positions
  of the objects O1 and O2 in S and S'
 corresponding to the 
  simultaneous observation of $E_1$ and $E_2$ in S, that yields the LFC. a) at the time
 of photon emission from O2, b) at $t = t' =0$, the time of observation in S of the LFC
  between $E_1$ and $E_2$ and c) at the time of photon emission from O1. The real
 positions of O1 and O2 are indicated by the cross hatched squares, and the emission (in S')
 or observation (in S) of photons by a star pattern.}}
\label{fig-fig5}
\end{center}
\end{figure}

 It is interesting to consider the sequence of space-time events, in the frames S, S', which 
 take place when the simultaneous observation is made in S of two objects, at rest in S', that
 yields the LFC. The correspondence, via the LT Eqns(3.1) and (3.2) of the relevant
  space-time events in S and S' is presented in Table 4 and shown in Fig.5.   
 \par When the mid-point of two objects  O1 and O2 considered in the previous Section
  (the origin, O', of S') passes
  the center of the barn (corresponding to the origin, O, of S)
   the clocks in the frames are synchronised such that $t = t' = 0$. The apparent and
  real positions of the objects, at this time, in S are as shown in Fig.5b.
   $E_1$ and $E_2$ represent
   space-time events coincident with the observation of single photons from O1 and O2
  respectively, that are both observed at $t = 0$ in S. According to the LT, the photon
   from O1 is emitted in S' at the later time $t' = \beta L/2c$ 
   (event $E_1'$ in Fig.5c), and that from O2 at the earlier time $t' = -\beta L/2c$
  (event $E_2'$ in Fig.5a), than the time ($t' = 0$) of observation of the photons in S.
  As before, $L$ is the separation of the objects in S'. The position of the origin of
   S' in S measures directly $t'$ via the relation, derived from Eqn(3.1),
   $ t'= x_{O'}^A/\gamma v = x_{O'}^R/\gamma v $. As can be seen in Fig.5
   and from the entries in Table 4, the real
   position of O' in S (mid-way between the real positions of O1 and O2) coincides with
   its apparent position, as calculated using Eqn(3.1), at all times.
   The real positions of O1 and O2 at the time, $t' = -\beta L/2c$,
  or $t = -\gamma \beta L/2c$, at which the observed photon (event $E_2$)
  is emitted from O2, are shown in Fig.5a.
   As above, $\gamma =2$, $\beta = \sqrt{3}/2$, $L = c = 1$ are assumed. Similarly Fig.5c shows 
  the real positions of the objects at the time $t' = \beta L/2c$, or $t = \gamma \beta L/2c$, 
  at which the observed photon (event $E_1$) is emitted from O1.
    It can be seen from Fig.5b that the apparent position of O2 in S, at $t = 0$, is shifted,
   relative to its real
  position, towards negative values of $x$, i.e. in the direction of the real
 position of O2 at the instant that the corresponding photon was emitted in S'.
 Similarly, the apparent position of O1 in S, at $t = 0$, is shifted towards
   positive values,
  i.e. also towards
 its real position in S at the time that the corresponding photon was emitted in S'.
 The combination of these shifts gives the LFC. It is a direct
 consequence of the relativity of simultaneity of the events observed in S and S'.
 However, the apparent positions of the objects in S do not correspond to the
 real positions of the objects at the times that the photons were emitted in S',
 as might naively be expected. In fact it can be seen from the entries in Table 4
 that at the time $t_2 = -\gamma \beta L/2c$, at which the photon observed at
 $E_2$ is emitted in S', the real position of O2 is:
 \begin{equation}
 x_{\rm{O2}}^R(t = t_2) = x_{\rm{O'}}^R(t = t_2)+\frac{L}{2} = \frac{L}{2}(1-\gamma \beta^2)
 \end{equation}
 which is separated from the apparent position of O2 at $t = 0$ by the distance:
  \begin{equation}
 x_{\rm{O2}}^A(t = 0) - x_{\rm{O2}}^R(t = t_2) = \frac{L}{2}(\frac{1}{\gamma}+\gamma \beta^2 -1)
  =  \frac{L}{2}(\gamma-1)
 \end{equation} The same distance separates the apparent position of 
  O1 at $t = 0$ from its real position at the time, $t_1 = \gamma \beta L/2c$,
  at which the photon was emitted from O1 in S'.      
 Table 4 and Fig.5 show that the photon from O2 is predicted by the LT
 to be observed
 in S at a time $\gamma \beta L/2c$ later than that corresponding
 to its emission time in S', whereas the photon from O1 is predicted to be
 observed in S at a time $\gamma \beta L/2c$ {\it before} O1 reaches the position
 at which the photon is emitted. The latter conclusion seems indeed paradoxical
 ~\cite{Soni} and merits further reflection. 
 \par A similar consideration of the Space Dilatation effect shows that this is also 
 follows directly from the relativity of simultaneity. Making a $\Delta t' = 0$
 projection in S' gives corresponding non-simultaneous events in S, spatially
 separated by the distance $\gamma L$ (see Table 1 of Reference~\cite{JHF2}).
\begin{table}
\begin{center}
\begin{tabular}{|c|c c c c c|} \hline
 Object
   & \multicolumn{1}{c|}{$x'$}
   & \multicolumn{1}{c|}{$t'$} 
   & \multicolumn{1}{c|}{ $x^A$}
   & \multicolumn{1}{c|}{ $x^R$}
   & \multicolumn{1}{c|}{ $t$} \\ \cline{1-6}
   &  &  &  &  &   \\   
 01 & $-\frac{L}{2}$ & $\frac{\beta L}{2c}$ & $-\frac{L}{2 \gamma}$ & $-\frac{L}{2}$ & 0 \\ 
   &  &  &  &  &   \\   
 0' &   0  & $\frac{\beta L}{2c}$ & $\frac{\gamma \beta^2 L}{2}$ & $\frac{\gamma \beta^2 L}{2}$ &
 $\frac{\gamma \beta L}{2c}$ \\
   &  &  &  &  & \\
 02 & $\frac{L}{2}$ & $-\frac{\beta L}{2c}$ & $\frac{L}{2 \gamma}$ &  $\frac{L}{2}$ & 0 \\ 
   &  &  &  &  &  \\   
 0' &   0  & $-\frac{\beta L}{2c}$ & $-\frac{\gamma \beta^2 L}{2 }$ & $-\frac{\gamma \beta^2 L}{2}$ &
 $-\frac{\gamma \beta L}{2c}$
 \\
   &  &  &  &  &  \\         
\hline
\end{tabular}
\caption[]{ { \sl Space-time points in S and S' of O1, O2 and O', at the times 
 $t'= \beta L/2c$ and $t'= -\beta L/2c$ of photon emission in S' corresponding to
 observation of the LFC effect in S at $t=0$. Real and apparent positions in
 S are denoted as $x^R$ and $x^A$ respectively. Note that there is distinction
  between real and apparent positions neither in S' nor, for the origin O', in S.}}      
\end{center}
\end{table}  

\SECTION{\bf{Summary and Conclusions}}
  
 The real positions of moving objects within a single frame of reference
 are calculated by the well-known rules, based on the concepts of spatial position,
 time, velocity and acceleration, as first clearly set down by Galileo. Special
 Relativity which instead relates, via the LT, space-time events in different
 inertial frames, has no relevance to the calculation of the relative positions
 of different moving objects in a single reference frame. The different apparent
 sizes of extended objects, or distances between discrete objects, predicted
 by the LT for different modes of observation, e.g. the LFC ($\Delta t = 0$
 projection) or SD ($\Delta t' = 0$  projection) manifest the truly apparent
  (i.e. observation mode dependent) nature of these phenomena.
 \par The incorrect conclusion concerning the Rockets-and-String paradox
  presented till now in the literature resulted from, on the one hand, a failure
  to distinguish between the (correctly calculated) real positions of the rockets
  and the apparent nature of the LFC, and on the other, by application of the LFC
  in an inconsistent manner, either to the string and not to the rocket separation
  in S, or to the rocket separation, but not to the string, in S'. There are no
  `stress effects' specific to SR and so the string does not break. 
  \par Similarly, all discussions of the Pole-and-Barn paradox (and several 
  similar examples cited above), in the literature and text books, have assumed
  that the LFC is a real, not an apparent, effect. There seems to be no
  justification for such an assumption in Einstein's original work on SR.
  When the real positions of the moving objects (or rod) in S are properly taken into 
  account it is evident that they can never actually fit into the barn so
  that the situation depicted in Figs.1 and 2 of Reference~\cite{Dewan} represents
  a physical absurdity.
  \par It is shown that both the LFC and SD effects are direct consequences of the
  relativity of simultaneity of SR first proposed by Einstein. In the case of the LFC,
  what seems to be a true paradox is revealed by this study. A photon, emitted by a moving
  object, is predicted, by the LT, to be observed before the moving object has reached the
  position at which it emits the  photon. Essentially the same apparent 
  breakdown  of causality as evidenced by `backwards running clocks'
  has been pointed out in a recently published paper~\cite{Soni}.

\pagebreak
 
 {\bf Acknowledgements} 
\par I would like to thank Y.Bernard for help in the preparation of the figures.

\pagebreak


\begin{thebibliography}{99}
\bibitem{Einstein}
A.Einstein,  
 Annalen der Physik {\bf17}, 891 (1905).
\bibitem{Pais}
 See the discussion of Einstein's
theory of Special Relativity, as presented in Reference[1] above, and related work
of Fitzgerald, Lorentz and Poincar\'{e} given in Chapters 7 and 8 of: A.Pais,
`Subtle is the Lord, the Science and Life of Albert Einstein',
Oxford University Press (1982).
\bibitem{Terrell}
J.Terrell,
Phys. Rev. {\bf 116},  1041  (1959).
\bibitem{Penrose}
R.Penrose,
Proc. Cambridge Phil. Soc. {\bf 55},  137 (1959).
\bibitem{Weinstein}
R.Weinstein,
Am. J. Phys. {\bf 28}, 607 (1960).
\bibitem{JHF1}
J.H.Field,`Space Time Measurements in Special Relativity' 
University of Geneva pre-print UGVA-DPNC 1998/04-176 April 1998,
physics/9902048.
Published in the Proceedings of the XX Workshop on High Energy
Physics and Field Theory, Protvino, Russia, June 24-26 1997.
Edited by I.V.Filimonova and V.A.Petrov pp214-248.
\bibitem{JHF2}
J.H.Field,
 Am. J. Phys. {\bf 68}, 367 (2000). 
\bibitem{Sorensen} 
R.A.Sorensen,
Am. J. Phys. {\bf 63}, 413 (1995).
\bibitem{Jefimenko}
O.D.Jefimenko,
Am. J. Phys. {\bf 64}, 812 (1996).
\bibitem{DewBer}
E.Dewan and M.Beran, Am. J. Phys. {\bf 27},  517 (1959).
\bibitem{Dewan}
E.Dewan,
  Am. J. Phys. {\bf 31},  383 (1963).
\bibitem{Rindler}
W.Rindler, Am. J. Phys. {\bf 29}, 365 (1961).
\bibitem{Shaw}
R.Shaw,  Am. J. Phys. {\bf 30},  72 (1962).
\bibitem{TayWheel}
 E.F.Taylor and J.A.Wheeler, `Spacetime Physics' (W.H.Freeman and
 Company, San Francisco, 1966).
\bibitem{TipLewell}
 P.A.Tipler and R.A.Lewellyn `Modern Physics' (W.H.Freeman and
 Company, New York, 2000. 
\bibitem{PerrJeff}
 This is the original 1923 translation of W.Perrett and G.B.Jeffrey, 
 later reprinted in `The Principle of Relativity', (Dover Publications,
 New York, 1952).
\bibitem{GermDict}
 H.Messinger, `Langenscheidts Grosses Schulw\"{o}rterbuch Deutsch-Englisch',
 (Langenscheidt KG, Berlin und M\"{u}nchen, 1977).
\bibitem{Marder}
 L.Marder `Time and the Space Traveller' (Allen and Unwin, London, 1971), Ch 3.  
\bibitem{Nikolic}
H.Nikolic, 
  Am. J. Phys. {\bf 67},  1007 (1999).
\bibitem{Rindler1}
W,Rindler,`Introduction to Special Relativity', 2nd Edition (O.U.P. Oxford, 1991)
 Section 14 P33.
\bibitem{Nawrocki}
P.J.Nawrocki,
  Am. J. Phys. {\bf 30},  771 (1962).
\bibitem{Romain} 
 J.E.Romain,
 Am. J. Phys. {\bf 31},  576 (1963).
\bibitem{EvWag} 
A.A.Evett and R.K.Wagness, 
  Am. J. Phys. {\bf 28}, 566  (1960).
\bibitem{Evett}
A.A.Evett,
 Am. J. Phys. {\bf 40},  1170 (1972).
\bibitem{TR}
 A.Tartaglia and M.L.Ruggiero, Eur. J. Phys. {\bf 24},  215 (2003).
\bibitem{Soni}
 It is essentially the same as the `backwards running clocks' 
  paradox \\
 noted recently. See  V.S.Soni, 
 Eur. J. Phys. {\bf 23}, 225 (2002).
\end{thebibliography}
\end{document}